\RequirePackage{fix-cm}
\documentclass[twocolumn,nofootinbib,showkeys]{revtex4-1}
\RequirePackage{graphicx}
\usepackage{mathbbol}
\usepackage{graphicx}
\usepackage{amsmath}
\begin{document}

\title{The exponential parameterization of the quark mixing matrix}


\author{G. Dattoli}
\email{giuseppe.dattoli@enea.it}

\author{E. Di Palma }
\email{emanuele.dipalma@enea.it}


\affiliation{
           ENEA - Centro Richerche Frascati, Via Enrico Fermi 45, 00044, Frascati, Rome, Italy}


\begin{abstract}
We comment on the exponential parameterization of the quark mixing matrix, by stressing that it naturally incorporates the Cabibbo structure and the hierarchical features of the Wolfenstein form. We extend our results to the neutrino mixing and introduce an exponential generator of the tribimaximal matrix.
\end{abstract}


\maketitle

\section{Introduction}
\label{intro}
The quark mixing matrix can be written in different ways, any of the proposed forms displays nice features and
disadvantages.
Whatever form one uses, four arbitrary parameters and the assumption of its unitarity are
necessary to get physically meaningful results. The models can be roughly grouped in two categories, the first
inspired to  Euler like rotation matrices, the second, containing explicit hierarchical features, employs an
 expansion, around the unit matrix, in term of some key parameters.
The original Kobayashi and Maskawa matrix\cite{Kobayashi} had been written in terms of three mixing angles $\theta_{1,2,3}$
and one CP violating phase $\delta$. In this parameterization the first family decouples from the others in the limit $\theta_1 \rightarrow 0$.
The particle data group\cite{Chau} chooses a form in which the CP violating term is appended to the matrix
entries responsible for the coupling of the first and third generations of quark mass eigenstates.
Finally Wolfenstein\cite{Wolfenstein} has proposed a matrix emerging from a kind of perturbative expansion
 in terms of the Cabibbo coupling parameter $\lambda\cong 0.22$\cite{Cabibbo}.
A third model, bridging between the (\cite{Kobayashi},\cite{Chau}) and \cite{Wolfenstein}, is based on the so called exponential parameterization,
which emerges from the request of unitarity, automatically satisfied by setting \cite{Dattoli}
\begin{equation}
\left. \begin{array}{ll}
        \widehat{V} &= e^{\widehat{A}}  \\
        \widehat{A}^{\dag} &= - \widehat{A}
     \end{array}
\right. \label{eq:equation1}
\end{equation}
The second condition in Eq. (\ref{eq:equation1}), expressing the anti-hermiticity of the matrix,
is ensured by the following specific choice
\begin{equation}
\widehat{A}=\left( \begin{array}{ccc}
        0 & \Lambda_1 & \Lambda_3  \\
        -\Lambda_1 &0 & \Lambda_2\\
        -\Lambda_3^* &-\Lambda_2 & 0\\
     \end{array}
\right) \label{eq:equation2}
\end{equation}
The vanishing of the diagonal entries secures that the matrix $\widehat{V}$ be unimodular\footnote{It would be sufficient to have a
matrix with null trace, but for practical reasons we use the form(2).}.
The sub-labels 1, 2, 3 determine the mixing d-s, s-b, d-b respectively, all the entries, except $\Lambda_3$, are real.
In the spirit of Wolfeinstein criteria, we use the Cabibbo strength $\lambda$
as key parameter and make the following identifications\cite{Dattoli2}
\begin{equation}
\left. \begin{array}{ll}
        \Lambda_1 &=\lambda  \\
        \Lambda_2  &= y \lambda^2 \\
        \Lambda_3  &= x\lambda^3 e^{\imath \delta}
     \end{array}
\right. \label{eq:equation3}
\end{equation}
containing an implicit hierarchical assumption on the coupling between the different quark families.
The vanishing of the $x$, $y$ coefficients allows the decoupling from the b sector reducing  the matrix
to the s-d Cabibbo mixing, namely
\begin{equation}
\widehat{V}_{(x,y)\rightarrow 0}=e^{
\left( \begin{array}{ccc}
        0 & \lambda & 0 \\
        -\lambda &0 & 0\\
        0 &0 & 0\\
     \end{array}
\right)}=
\left( \begin{array}{ccc}
        \cos(\lambda) & \sin(\lambda) & 0 \\
        -\sin(\lambda) &\cos(\lambda) & 0\\
        0 &0 &1\\
     \end{array}
\right) \label{eq:equation4}
\end{equation}
It is also to be stressed that $\widehat{V}_{\lambda \rightarrow 0}= \widehat{\mathbb{1}}$, which means that the parameterization in (\ref{eq:equation3}) contains the assumption
that the vanishing of the Cabibbo parameter determines the decoupling of the entire quark matrix.
The phase $\delta$ is associated, as in the particle data group choice, with the smallest coupling term.
In this paper we will see how the exponential parameterization yields a flexible tool to analyse
the quark mixing phenomenology and the relevant consequences.

\section{The matrix A and the Wolfenstein parameterization}
We will prove that the quark mixing matrix written as in Eq. (\ref{eq:equation1}) naturally contains
the Wolfenstein parameterization and the Euler like forms as well.
By keeping the expansion of the exponential in Eq. (\ref{eq:equation1}) up to third order in $\lambda$, namely
\begin{equation}
\widehat{V}=\widehat{\mathbb{1}}+\widehat{A}+\frac{\widehat{A}^2}{2}+\frac{\widehat{A}^3}{3!}+\frac{\widehat{A}^4}{4!}+o(\lambda^5)
\label{eq:equation5}
\end{equation}
we obtain the mixing matrix in the form
\begin{equation}
\begin{array}{l}
\widehat{V}\cong
\left( \begin{array}{ccc}
        1-\frac{\lambda^2}{2}+\frac{\lambda^4}{4!} & \lambda-\frac{\lambda^3}{3!} & A F \lambda^3 \\
        -\lambda+\frac{\lambda^3}{3!} & 1-\frac{\lambda^2}{2}+\frac{\lambda^4}{4!}-\frac{(A\lambda^2)^2}{2} &\frac{ A B \lambda^2}{2}\\
        A G \lambda^3  & \frac{A C \lambda^2}{2}  &1-\frac{A^2 \lambda^4}{2} \\
     \end{array}
\right) \\ \\
 A=y, \;F=\rho-\imath \eta,\; G=1-\rho-\imath \eta  \\
 B=2-\lambda^2(\rho-\frac{1}{6}-\imath \eta),\; B+C=-2\lambda^2(\rho-\frac{1}{2})\\
 \rho=\frac{x}{y} \cos(\delta)+\frac{1}{2},\; \eta=-\frac{x}{y}\sin(\delta)
 \end{array}
   \label{eq:equation6}
\end{equation}
Eq. (\ref{eq:equation6}) is recognized as a Wolfenstein-type parameterization, the Taylor expansion at higher order
can provide more accurate expansion in the Cabibbo coupling parameter, as we will see in the following.
The expansion at the third order allows a one to one correspondence between the Wolfenstein
parameters and those of the matrix $\widehat{A}$, which can be written in the form
\begin{equation}
\widehat{A}=\left( \begin{array}{ccc}
        0 & \lambda & A\lambda^3(\rho-\imath \eta-\frac{1}{2})  \\
        -\lambda & 0 & A \lambda^2\\
        -A\lambda^3(\rho+\imath \eta-\frac{1}{2})&-A \lambda^2 & 0
     \end{array}
\right) \label{eq:equation7}
\end{equation}
Using for $A,\; \rho, \; \eta$ the following values, close to those given in the literature \cite{DataGroup}:
\[
\begin{array}{ll}
\lambda =  0.2272\pm 0.0010 , &  \;\;\;\;\; A= 0.818^{+0.007}_{-0.017} \\ \\
 \rho= 0.221^{+0.064}_{-0.028},&  \;\;\;\;\; \eta= 0.340^{+0.017}_{-0.045}
\end{array}
\]
we find for $x$ and $\delta$ the following values
\[
x= -0.359^{+0.049}_{-0.052} , \;\;\; \delta= 0.883^{+0.145}_{-0.118}
\]
and we get for the mixing matrix\footnote{ This result has been obtained by
expanding the matrix at any arbitrary order, namely $\widehat{V}=\displaystyle{\sum_0^N} \frac{A^n}{n!}$ and by keeping N=50.
We have not included the errors deriving from the experimental and systematic uncertainties,
the relevant analysis would require extreme care for fitting the data and such an effort is
out of the purposes of the present note.}
\begin{equation}
|A|=\left( \begin{array}{ccc}
        0.97429 & 0.22523 & 3.86\cdot10^{-3}  \\
        0.22512 & 0.97341 & 0.04215 \\
        8.10 \cdot 10^{-3} & 0.04154 & 0.99910 \\
     \end{array}
\right) \label{eq:equation8}
\end{equation}

in good agreement with the values reported in\cite{DataGroup}. Higher orders expansions will be
considered in the forthcoming sections.

\section{The geometrical meaning of the exponential parameterization and the Euler like forms}

\begin{figure}[h]
\includegraphics[scale=0.4]{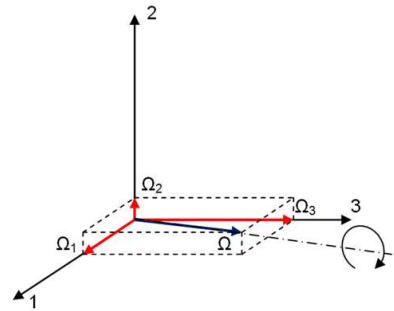}
\caption{The Quark mixing torque vector}
\label{fig:fig1}       
\end{figure}

\begin{figure}[h]
\includegraphics[scale=0.53]{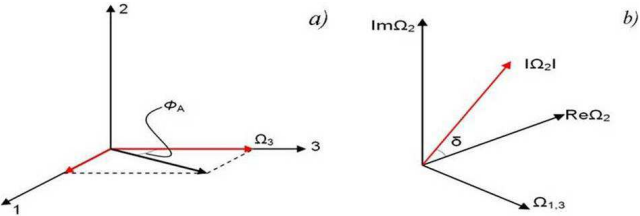}
\caption{The real(a) and the imaginary part (b) of the torque vector \mbox{$\Omega$}.}
\label{fig:fig2}       
\end{figure}

We have so far proved that the exponential parameterization of the mixing matrix has some nice
features which makes its use quite interesting.
Before going further let us speculate on the geometrical (physical) meaning of the matrix $\widehat{A}$,
which can be understood as a kind of Hamiltonian ruling the process of quark mixing.
We introduce, therefore, the Schroedinger equation
\begin{equation}
\imath \partial_{\tau}\underline{\psi}=\widehat{H}\underline{\psi}
\label{eq:equation9}
\end{equation}
where $\underline{\psi}|_{\tau=0}$ are the quark mass eigenstates, and
\begin{equation}
\widehat{H}\propto \imath \widehat{A}
\label{eq:equation10}
\end{equation}
Within such a picture the matrix $\widehat{V}$ is the evolution operator associated with Eq. (\ref{eq:equation9}).
In the case of vanishing CP phase $\delta\rightarrow 0$, the Hamiltonian in (\ref{eq:equation10}) can be written
in terms of SO(3) generators, namely
\begin{equation}
\widehat{H}=\lambda R_1+y\lambda^2 R_2+x\lambda^3 R_3
\label{eq:equation11}
\end{equation}
with
\[
\begin{array}{l}
R_1=\imath \left( \begin{array}{ccc}
        0 & 1 & 0  \\
        -1 & 0 & 0\\
        0 & 0 & 0\\
     \end{array}
\right) ,\;
R_2=\imath \left( \begin{array}{ccc}
        0 & 0 & 0  \\
        0 & 0 & 1\\
        0 & -1 & 0\\
     \end{array}
\right)  \\ \\
R_3=\imath \left( \begin{array}{ccc}
        0 & 0 & 1  \\
        0 & 0 & 0\\
        -1 & 0 & 0\\
     \end{array}
\right)
\end{array}
\]
The Schroedinger equation (\ref{eq:equation9}) can, accordingly, be viewed as a vector equation of the type
\begin{equation}
\left. \begin{array}{ll}
        \partial_{\tau} \overrightarrow{Q} =& \overrightarrow{\Omega} \times \overrightarrow{Q}  \\ \\
        \overrightarrow{\Omega} \equiv & \lambda(-y\lambda,x\lambda^2,-1)
     \end{array}
\right. \label{eq:equation12}
\end{equation}
where $\overrightarrow{Q}\equiv(\psi_1,\psi_2,\psi_3)$ is the vector associated with the quark field.
The problem of the quark mixing is therefore understood as a rotation, induced by an Euler-like torque equation. \\
The torque vector $\overrightarrow{\Omega}$ is reported in Fig. (\ref{fig:fig1}) along with the role played by each vector component.
The quark mixing matrix can be written using the Cayley Hamilton theorem \cite{Babusci} (see Sect.~\ref{sec:4}) as
\begin{equation}
\left. \begin{array}{l}
        \widehat{V}=e^{\tau \widehat{A}}|_{\tau=1}=\widehat{\mathbb{1}}+\mbox{sinc}(|\overrightarrow{\Omega}|)\widehat{A}+\displaystyle{ \frac{1}{2}(\mbox{sinc}(\frac{|\overrightarrow{\Omega}|}{2}))^2 \widehat{A}^2}  \\ \\
        |\overrightarrow{\Omega}|=\lambda\sqrt{1+y^2 \lambda^2+x^2 \lambda^4}\\ \\
        \mbox{sinc}(\alpha)=\displaystyle \frac{\sin(\alpha)}{\alpha}\\
     \end{array}
\right. \label{eq:equation13}
\end{equation}
Moreover from Eq. (\ref{eq:equation12}) the action of the mixing matrix on the initial vector $\overrightarrow{Q}$ can be specified through the following Rodriguez rotation\cite{Babusci1}
\begin{equation}
\left. \begin{array}{rl}
        \overrightarrow{Q}=&\cos(|\overrightarrow{\Omega}|)\overrightarrow{Q}_0+\sin(|\overrightarrow{\Omega}|) \overrightarrow{n}\times \overrightarrow{Q}_0 \\ \\
        &+(1-\cos(|\overrightarrow{\Omega}|))(\overrightarrow{n} \cdot \overrightarrow{Q}_0)\cdot \overrightarrow{n}\\ \\
        \overrightarrow{n}=&\frac{\overrightarrow{\Omega}}{|\overrightarrow{\Omega}|}
     \end{array}
\right. \label{eq:equation14}
\end{equation}
The geometrical interpretation is less obvious if we include the CP violating term.
We assume Eq. (\ref{eq:equation11}) to be still valid and with a slight abuse of the notation write
\begin{equation}
\left. \begin{array}{rll}
        \overrightarrow{\Omega}   &\equiv& \overrightarrow{\Omega}_1+\imath \overrightarrow{\Omega}_2 \\ \\
        \overrightarrow{\Omega}_1 &\equiv& (-y\lambda^2,x\lambda^3 \cos(\delta),-\lambda)\\ \\
        \overrightarrow{\Omega}_2 &\equiv& (0,x\lambda^3\sin(\delta),0)
     \end{array}
\right. \label{eq:equation15}
\end{equation}
This assumption contains the bare essence of CP violation from a geometrical point of view.
The vector $\overrightarrow{\Omega}$ splits into a real and imaginary part, as shown in Fig. (\ref{fig:fig2})
where the second component of the torque vector is composed by two subcomponents:
\begin{description}
  \item[a)]  the coupling vector $\overrightarrow{\Omega}_{1,3}\equiv(y\lambda^2,0,-\lambda)$  is the component of the vector   in the 1-3 plane;
  \item[b)]  the CP violating sector is viewed as the pseudo vector $(\Omega_{1,3},\mbox{Im}(\Omega_2),\mbox{Re}(\Omega_2))$.
\end{description}

In terms of the Wolfenstein parameters the modulus of the Torque vector can be written as
\begin{equation}
 |\overrightarrow{\Omega}|=\sqrt{\lambda^2+(A\lambda^2)^2+\left[\left(\rho-\frac{1}{2} \right)^2+\eta^2  \right](A\lambda^3)^2}
 \label{eq:equation16}
\end{equation}
or as

\begin{equation}
\left. \begin{array}{l}
        |\overrightarrow{\Omega}|=\sqrt{\lambda^2(1+\tan^2(\phi_A))+\lambda^3\left(\rho-\frac{1}{2} \right)^2} \\ \\
           \;\;\;\;\;\;\;\;\;\; \;\;\;\;\;\;\;\;\;\; \;\;\;\;\;\;\;\;\;\; \;\;\;\;\;\;\;\ \overline{\left[1+\tan^2(\delta)  \right]\tan^2(\phi_A)} \\ \\
        \tan(\phi_A)=A\lambda, \;\;\; \tan(\delta)=\displaystyle{\frac{\eta}{\sqrt{\left(\rho-\frac{1}{2}\right)^2}}}
     \end{array}
\right. \label{eq:equation17}
\end{equation}

where $\phi_A$ and $\delta$ are indicated in Figs. (\ref{fig:fig2}).\\
The angle $\phi_A$ lies in the (1,3) sector and specifies the direction of the
$\overrightarrow{\Omega}$ vector components in this plane. We visualize the geometric content of our problem
as indicated in the second of Figs. (\ref{fig:fig2}), in which the complex vector
component lying along the direction of the axis 2 is split into an imaginary and a real part. \\
In more rigorous mathematical terms we can illustrate the above procedure as it follows.
We first note that
\begin{equation}
\widehat{A}=\widehat{A}_1+\widehat{A}_2
\label{eq:equation18}
\end{equation}
with
\[
\widehat{A}_2=\left( \begin{array}{ccc}
       0 & 0 & \Lambda_3\\
       0 & 0 & 0\\
       -\Lambda_3^{*} & 0 & 0
       \end{array}
       \right), \;\;
     \widehat{A}_1=
     \left( \begin{array}{ccc}
       0 & \Lambda_1 & 0\\
       -\Lambda_1 & 0 & \Lambda_2\\
      0 & -\Lambda_2 & 0
       \end{array}
       \right)
\]

The matrices labelled with 2, 1 are not commuting each other, therefore we have at the first order in the
Zassenhaus disentanglement formula\footnote{ The Zassenhaus formula writes
$e^{\widehat{A}+\widehat{B}}=e^{\widehat{A}}e^{\widehat{B}}\displaystyle{\prod_{n=1}^{\infty}}e^{\widehat{C}_m}$ where the operators
$\widehat{C}_m$ are given in terms of successive commutators, the first two being
$\widehat{C}_1=-\frac{1}{2}[\widehat{A},\widehat{B}]$,
$\widehat{C}_2=\frac{1}{3}[\widehat{A},\widehat{B}]+\frac{1}{6}[\widehat{A},[\widehat{A},\widehat{B}]]$} \cite{Magnus}

\begin{equation}
\left. \begin{array}{l}
        \widehat{V}=e^{\widehat{A}_2+\widehat{A}_1}\cong e^{\widehat{A}_2}e^{\widehat{A}_1}e^{\widehat{C}} \\ \\
        \widehat{C}=-\frac{1}{2}[\widehat{A}_2,\widehat{A}_1]
     \end{array}
\right. \label{eq:equation19}
\end{equation}
Where the (anti-hermitian) matrix  $\widehat{C}$ is given by
\begin{equation}
\begin{array}{rcl}
\widehat{C}&=&-\frac{1}{2}\left( \begin{array}{ccc}
        0 & -\Lambda_3 \Lambda_2& 0  \\
        \Lambda_3^* \Lambda_2 & 0 & \Lambda_1 \Lambda_3\\
        0&-\Lambda_1 \Lambda_3^* & 0\\
     \end{array}
\right)= \\ \\
&=&-\displaystyle{\frac{\lambda^4 x}{2}}
\left( \begin{array}{ccc}
        0 & - \lambda y e^{\imath \delta}& 0  \\
        \lambda y e^{-\imath \delta} & 0 & e^{\imath \delta}\\
        0&e^{-\imath \delta} & 0\\
     \end{array}
\right)
\end{array}
 \label{eq:equation20}
\end{equation}
Neglecting the matrix $\widehat{C}$, which is of the order $o(\lambda^4)$,  we find that the CKM matrix can be expressed as
\begin{equation}
\widehat{V}\cong e^{\widehat{A}_2}e^{\widehat{A}_1}
\label{eq:equation21}
\end{equation}
With
\begin{equation}
e^{\widehat{A}_2}=\widehat{V}_2=\left( \begin{array}{ccc}
        \cos(|\Lambda_3|) & 0& \frac{\Lambda_3}{|\Lambda_3|}\sin(|\Lambda_3|)  \\
        0 & 1 & 0\\
        -\frac{\Lambda_3^*}{|\Lambda_3|}\sin(|\Lambda_3|)&0 & \cos(|\Lambda_3|)\\
     \end{array}
\right)
 \label{eq:equation22}
\end{equation}
and the use of the Cayley Hamilton theorem allows the following (exact) form of the second exponential
\[
e^{\widehat{A}_1}=\widehat{V}_1=C_0 \widehat{\mathbb{1}}+C_1 \widehat{A}_1+C_2
\widehat{A}_1^2
\]
\begin{equation}
\left( \begin{array}{c}
        C_0  \\ \\
        C_1\\ \\
        C_2\\
     \end{array}
\right)=
\left( \begin{array}{ccc}
        |\Lambda_{1,2}|^2 & 0 & 0  \\ \\
        -\Upsilon & \Upsilon & \Upsilon\\ \\
        \imath&1-\imath & -(1+\imath)\\
     \end{array}
\right)\left( \begin{array}{c}
        \displaystyle{\frac{1}{|\Lambda_{1,2}|^2}}  \\ \\
        \displaystyle{\frac{e^{\imath  |\Lambda_{1,2}|}}{2|\Lambda_{1,2}|^2}}\\ \\
       \displaystyle{\frac{e^{-\imath  |\Lambda_{1,2}}| }{2|\Lambda_{1,2}|^2}}\\
     \end{array}
\right)
 \label{eq:equation23}
\end{equation}

\[
\Upsilon=|\Lambda_{1,2}| (1+\imath), \;\;
|\Lambda_{1,2}|=\sqrt{\Lambda_1^2+\Lambda_2^2}
\]
The above formulae are a restatement of the tentative geometrical picture of  Fig. (\ref{fig:fig2}).
The naïve disentanglement has reduced the CKM generation to the product of two matrices,
$\widehat{V}_1$ accounting for the mixing, induced by the vector $\overrightarrow{\Omega}_{1,3}$ ,
and $\widehat{V}_2$ specifying a complex rotation, responsible for the CP violating contributions. \\
The matrix (\ref{eq:equation21}) is an approximation of the exponential form at the order $o(\lambda^4)$,
but it is not equivalent to Wolfenstein matrix. The matrix (\ref{eq:equation21}),
albeit an approximation, since we have neglected higher order commutators,
is unitary at any order in the coupling parameter, while
$\widehat{V}_W \widehat{V}_W^{\dag}=\widehat{\mathbb{1}}+o(\lambda^4)$ (where $\widehat{V}_W$
is the matrix (\ref{eq:equation6})). \\
We have stressed that the simple picture in terms of  Euler rotation is hampered
by the presence of a complex term, the $\widehat{V}$ matrix cannot be written in terms of
the  generators of rotations and indeed we find
\begin{equation}
\begin{array}{rl}
\widehat{V}=& e^{-\imath(\lambda \widehat{R}_1+y \lambda^2 \widehat{R}_2+x \lambda^3 \widehat{T})} \\ \\

\widehat{T}=&-\imath \left( \begin{array}{ccc}
        0 & 0& e^{\imath \delta}  \\
         & 0 & 0\\
        -e^{-\imath \delta}&0 & 0\\
     \end{array}
      \right)
 \end{array}\label{eq:equation24}
\end{equation}
The $\widehat{T}$ matrix does not belong to SO(3) and the quark mixing matrix,
written as the product of the exponential matrix correct up to the order $o(\lambda^4)$ is
\[
\widehat{V}=e^{- \imath x \lambda^3 \widehat{T}}\;
e^{- \imath y \lambda^2 \widehat{R}_2}\; e^{- \imath \lambda^3 \widehat{R}_1}+o(\lambda^4)
\]

\begin{equation}
\begin{array}{rl}
\widehat{V} \cong& e^{ \left( \begin{array}{ccc}
                            0 & 0& x \lambda^3 e^{\imath \delta}  \\
                            0& 0 & 0\\
                            -x \lambda^3 e^{-\imath \delta}&0 & 0\\
                        \end{array}
                    \right)} \\ \\
                 &\;\;\;\;\; e^{ \left( \begin{array}{ccc}
                    0 & 0& 0 \\
                    0 & 0 & y \lambda^2\\
                    0&-y \lambda^2 & 0\\
                    \end{array}
                    \right)}
                        e^{ \left( \begin{array}{ccc}
                        0 & \lambda & 0 \\
                        -\lambda & 0 & 0\\
                        0&0 & 0\\
                    \end{array}
                \right)}= \\ \\
               =&
                { \left( \begin{array}{ccc}
                            C(x\lambda^3) & 0&  e^{\imath \delta} S(x \lambda^3) \\
                            0& 1 & 0\\
                            -e^{-\imath \delta} S(x \lambda^3)&0 & C(x\lambda^3)\\
                        \end{array}
                    \right)} \\ \\
                &\;\;\;     { \left( \begin{array}{ccc}
                    1 & 0& 0 \\
                    0 & C(y\lambda^2) & S(y \lambda^2)\\
                    0&-S(y \lambda^2) & C(y\lambda^2)\\
                    \end{array}
                    \right)}
                        { \left( \begin{array}{ccc}
                        C(\lambda) & S(\lambda) & 0 \\
                        -S(\lambda) & C(\lambda) & 0\\
                        0&0 & 1\\
                    \end{array}
                \right)} \\ \\
              & C(\phi)=\cos(\phi), \; S(\phi)=\sin(\phi)
\end{array}
\label{eq:equation25}
\end{equation}
and displays the largely well-known feature that the mixing
angles are proportional to the Cabibbo coupling parameter according to
\begin{equation}
\left. \begin{array}{ccc}
        \vartheta_{1,3}& \propto & \lambda^3\cong  \displaystyle{ \sqrt{\frac{m_d}{m_b}}}\\ \\
        \vartheta_{2,3}&\propto& \lambda^2\cong \displaystyle{ \sqrt{\frac{m_s}{m_b}}}\\ \\
        \vartheta_{1,2}&\propto& \lambda\cong \displaystyle{\sqrt{\frac{m_d}{m_s}}}\\ \\
     \end{array}
\right. \label{eq:equation26}
\end{equation}

Furthermore, in full agreement with the particle data group paradigm, we get
\begin{equation}
\begin{array}{l}
e^{ \left( \begin{array}{ccc}
                            0 & 0& x \lambda^3 e^{\imath \delta}  \\
                            0& 0 & 0\\
                            -x \lambda^3 e^{-\imath \delta}&0 & 0\\
                        \end{array}
                    \right)}=
         \widehat{U}_{\delta}^{\dag}e^{-\imath x \lambda^3 \widehat{R}_3}\widehat{U}_{\delta} \\ \\
\widehat{U}_{\delta}=e^{\imath \widehat{\delta}}, \;\; \mbox{with}\;\;
     \widehat{\delta}= \left( \begin{array}{ccc}
                            0 & 0& 0  \\
                            0& 0 & 0\\
                            0&0 & \delta\\
                        \end{array}
                    \right), \;
               \widehat{R}_3= \imath \left( \begin{array}{ccc}
                            0 & 0 & 1 \\
                            0 & 0 & 0\\
                           -1 & 0 & 0\\
                        \end{array}
                    \right)
\end{array}
\label{eq:equation27}
\end{equation}
It is also interesting to note that the $\widehat{T}$ matrix can be written as

\begin{equation}
\begin{array}{l}
\widehat{T}=-\imath \cos(\delta) \widehat{R}_3+\imath \sin(\delta) \widehat{S}_3 \\ \\
    \widehat{S}_3=\left( \begin{array}{ccc}
                    0 & 0& 1  \\
                    0 &0 & 0\\
                    1&0 & 0 \\
                    \end{array}
                    \right)
\end{array} \label{eq:equation28}
\end{equation}
The naïve disentanglement (order $o(\lambda^4)$)
\begin{equation}
\begin{array}{l}
\widehat{V}\cong e^{-\imath \widehat{R}} e^{ \imath x \lambda^3  \sin(\delta) \widehat{S}_3}\\ \\
\widehat{R}=\lambda \widehat{R}_1+y \lambda^2 \widehat{R}_2+ x\lambda^3 \cos(\delta)\widehat{R}_3
\end{array}
\end{equation}

Corresponds to the product of two matrices, namely
\begin{equation}
\begin{array}{l}
\widehat{V} \cong \widehat{V}_R \widehat{V}_I \\ \\
e^{\widehat{R}}=\widehat{V}_R \\ \\
    \widehat{V}_I=\left( \begin{array}{ccc}
                    C(x\lambda^3 \sin(\delta)) & 0& \imath S(x\lambda^3 \sin(\delta)) \\
                    0 &1 & 0\\
                    \imath S(x\lambda^3 \sin(\delta))&0 & C(x\lambda^3 \sin(\delta)) \\
                    \end{array}
                    \right) = \\ \\
 \;\;\;\; =\left( \begin{array}{ccc}
                    C(A \lambda^3 \eta) & 0& -\imath S( A \lambda^3 \eta) \\
                    0 &1 & 0\\
                    -\imath S(A \lambda^3 \eta)&0 & C( A\lambda^3 \eta) \\
                    \end{array}
                    \right)
\end{array} \label{eq:equation29}
\end{equation}
and $\widehat{V}_R$ can be written as Eq. \ref{eq:equation13} with
\begin{equation}
\overrightarrow{\Omega}\equiv \lambda(-y \lambda,x \lambda^2 \cos(\delta),-1)
\end{equation}

The matrix $\widehat{V}_I$ mixes the first and third quark generation mass eigenstates and is responsible
for the CP violation. It is a pseudo rotation matrix and is generated by a matrix whose
determinant is the Jarlskog invariant \cite{Jarlskog}, discussed in the forthcoming section. \\
We have so far shown that the exponential parameterization implicitly contains
Wolfenstein and Euler type forms, in the following sections we will dwell on its further advantages.

\section{The Cayley Hamilton Theorem and The Quark mixing matrix}
\label{sec:4}
The exponential matrix (\ref{eq:equation1}) can be treated in different ways. \\
We have already shown that the use of  a Taylor expansion leads to a Wolfenstein form,
which preserves the unitarity of $\widehat{V}$ at the expansion order (the mixing matrix in Eq. (\ref{eq:equation6})
is unitary at the order $o(\lambda^4)$). \\
The method of the exponential disentanglement can be used too and such a procedure allows an interesting
geometrical picture of the mixing dynamics and albeit an approximation in the Cabibbo coupling parameter,
the mixing matrix written as in Eq. (\ref{eq:equation17}) preserves the unitarity at any order in $\lambda$,
as discussed more accurately in the concluding remarks. \\
The matrix $\widehat{V}$ can, however, be written in an exact form using the Cayley Hamilton theorem,
by setting
\begin{equation}
\widehat{V}=C_0 \widehat{\mathbb{1}}+C_1 \widehat{A}+C_2 \widehat{A}^2
\label{eq:equation31}
\end{equation}
where
\begin{equation}
        e^{\varepsilon_j}=C_0 +  \varepsilon_j C_1 + \varepsilon_j^2 C_2, \;\; \mbox{with}\;\;  j=1,2,3
\end{equation}
With $\varepsilon_j$  being the roots associated with the characteristic equation of the matrix $\widehat{A}$,
namely
\begin{equation}
       \varepsilon_j^3+|\overrightarrow{\Omega}|^2 \varepsilon_j+ \imath \triangle =0
        \label{eq:equation33}
\end{equation}
where
\begin{equation}
\left.
     \begin{array}{l}
       \triangle=2 xy \lambda^6 \sin(\delta)=-2 A \eta \lambda^6 \\ \\
        |\overrightarrow{\Omega}|=\lambda\sqrt{1+y^2\lambda^2+ x^2\lambda^4}= \\ \\
     =\lambda\sqrt{1+(A \lambda)^2+(A^2 \lambda^4)\left[\left(\rho-\frac{1}{2} \right)^2+\eta^2 \right]}
     \end{array}
\right.
\end{equation}
$\imath\triangle$ is the determinant of the matrix $\widehat{A}$.\\
A little bit of algebra yields to define
the $C_i\;\;(i=0,1,2)$ coefficients as the product of two matrix
\begin{multline}
\left(
     \begin{array}{c}
     C_0 \\ \\
     C_1 \\ \\
     C_2
     \end{array}
\right)= \left( \begin{array}{ccc}
                    \varepsilon_2 \varepsilon_3 & \varepsilon_1 \varepsilon_3& \varepsilon_1 \varepsilon_2 \\ \\
                    -(\varepsilon_2 + \varepsilon_3) &-(\varepsilon_1 + \varepsilon_3) & -(\varepsilon_1 +\varepsilon_2)\\ \\
                    1&1 & 1
                    \end{array}
                    \right) \\
            \left(
                \begin{array}{c}
                \displaystyle{\frac{e^{\varepsilon_1}}{(\varepsilon_2 -\varepsilon_1)(\varepsilon_3 -\varepsilon_1)} }\\ \\
                \displaystyle{\frac{e^{\varepsilon_2}}{(\varepsilon_1 -\varepsilon_2)(\varepsilon_3 -\varepsilon_2)}} \\ \\
               \displaystyle{\frac{e^{\varepsilon_3}}{(\varepsilon_1 -\varepsilon_3)(\varepsilon_2 -\varepsilon_3)}}
                \end{array}
            \right)
\end{multline}
Eq. (\ref{eq:equation23}) (along with Eqs. (\ref{eq:equation27})) is the most general form of
the quark mixing matrix which can be derived from an exponential parameterization, it is exact but not easy to remember. \\
Let us now give an idea of the orders of the numerical values characterizing the various
quantities entering the above equations.
The use of the previously quoted values for the Wolfenstein parameters lead to the following evaluations for the solution of Eq. (\ref{eq:equation33})
\begin{equation}
\begin{array}{l}
\varepsilon_1 \cong  -0.23171  \imath \\ \\
\varepsilon_2 \cong  0.00117 \imath \cdot  \\ \\
\varepsilon_2 \cong  0.23054 \imath \cdot
\end{array}
\end{equation}
It is worth stressing that the matrix $\widehat{D}$ provides the diagonal forms of either
$\widehat{V}$ and $\widehat{A}$. It follows therefore that the two matrices have the same eigenvectors.
They can be determined using $\widehat{A}$ instead of $\widehat{V}$,
because the procedure is significantly simpler. We find  that the eigenvalues are in the form
\begin{equation}
|j>=\left( \begin{array}{c}
        1\\ \\
        -\varepsilon_j-x y \lambda^5 e^{-\imath \delta}\\ \\
        -y \lambda^3 - \varepsilon_j x \lambda^3 e^{-\imath \delta}\\ \\
     \end{array}
\right)
\end{equation}

\begin{table*}
\begin{multline}
\widehat{V}=\left( \begin{array}{cc}
        C(\lambda)+\frac{A^2 \lambda^6}{4!}\Phi & S(\lambda)-\frac{A^2 \lambda^5}{2}(\Pi^*-\frac{1}{6})  \\ \\
       -S(\lambda)-\frac{A^2 \lambda^5}{2}(\Pi-\frac{5}{6}) &  C(\lambda)-(A\lambda^2)^2 \left[ \frac{1}{2}-\frac{\lambda^2}{3}\left( \frac{1}{4}-\imath \eta \right)\right] \\ \\
        -A\lambda^3\left[ \left(\Pi-1\right)-\frac{\lambda^2}{6}\left( \Pi-\frac{3}{4} \right) \right] &
                         -S(A\lambda^2)+\frac{A \lambda^4}{4!} \left( \lambda^2(\Pi-\frac{7}{10})-12(\Pi- \frac{2}{3})  \right)
     \end{array}
\right.  \\
\left. \begin{array}{c}
          A\lambda^3\left[ \Pi^*-\frac{\lambda^2}{6}\left( \Pi^* -\frac{1}{4} \right) \right]\\ \\
          S(A\lambda^2) + \frac{A \lambda^4}{4!} \left( \lambda^2(\Pi^*-\frac{3}{10})-12(\Pi^*- \frac{1}{3}) \right)\\ \\
          1-\frac{(A \lambda^2)^2}{2}+\frac{A^2 \lambda^6}{4!}\Phi\\ \\
     \end{array}
\right)
\label{eq:equation42}
\end{multline}
\[
\begin{array}{l}
\Phi=-12(\rho^2+\eta^2)-8\imath \eta+12 \rho-2; \;\; \Pi=\rho+\imath \eta; \;\;  \Pi^*=\rho-\imath \eta
\end{array}
\]
\end{table*}

It is worth mentioning the companion matrix associated with the
characteristic equation (30) \cite{Companionmatrix}, which writes
\begin{equation}
C_A=\left( \begin{array}{ccc}
        0\;\;&\;\; 0 & -\varepsilon_1 \varepsilon_2 \varepsilon_3\\ \\
        1\;\;&\;\; 0 & -(\varepsilon_1 \varepsilon_2+\varepsilon_2 \varepsilon_3 +\varepsilon_1 \varepsilon_3)\\ \\
        0\;\;&\;\;1& -(\varepsilon_1 + \varepsilon_2 + \varepsilon_3)\\ \\
     \end{array}
\right) \label{eq:equation40}
\end{equation}
It is accordingly expressed in terms of three invariants\footnote{A $3\times3$ matrix has three invariants
given by its determinant, its trace and by the sum of the determinants of its minors}, namely
\begin{equation}
\left.
     \begin{array}{l}
         \displaystyle \varepsilon_1 \varepsilon_2 \varepsilon_3= -2 \imath x \lambda^6 \sin(\delta)\\ \\
         \varepsilon_1^2 + \varepsilon_2^2+ \varepsilon_3^2= |\overrightarrow{\Omega}|^2 \\ \\
        \varepsilon_1+ \varepsilon_2 +\varepsilon_3= 0
     \end{array}
\right.
\end{equation}
the first of which is the Jarlskog invariant, a measure
of the amount of CP violations, emerging in quite a natural way in the present analysis.

\section{Concluding remarks}
We have shown that the exponential parameterization interpolates between Wolfestein
and Euler like forms and could provide a useful and flexible tool of analysis.
Its approximations in terms of the Cabibbo coupling can be either expressed as Taylor
expansions or as unitarity preserving forms bassed on the Zassenhaus formula. \\
The Taylor expansion does not meet too much aesthetical criteria, but it can usefully be
exploited to get higher order approximations of Wolfenstein type parameterizations an example is
shown below, where we report the naïve expansion of the exponential matrix up to the order $o(\lambda^7)$.
\\We have reported the matrix (\ref{eq:equation42}) (where $C(\lambda),\; S(\lambda)$ denote the expansion of cosine and sine up to the order $o(\lambda^7)$) for comparison purposes with other forms available in literature.
The accuracy of this last matrix is one part over $10^9$ and can therefore considered exact for any expansion
purposes. \\
The extension of the CKM matrix to higher dimensions by the use of the exponential
matrix method is not complicated. In the case of  four quark generations,
we define the matrix containing 2 CP violating phases, appended to the smallest
coupling terms. We have furthermore assumed that the coupling strengths to the fourth
family be of the order $\lambda^{3+n},\;\; n=1,2,3$.
\begin{equation}
  A=\left( \begin{array}{cccc}
                    0 & \lambda& e^{\imath \delta_1} x \lambda^3 &e^{\imath \delta_2} z \lambda^6  \\
                    -\lambda &0 & y\lambda^2  & p\lambda^5   \\
                    -e^{-\imath \delta_1} x \lambda^3 &  -y\lambda^2 & 0   & u\lambda^4\\
                    -e^{-\imath \delta_2} z \lambda^6 & -p\lambda^5 & -u\lambda^4& 0
                    \end{array}
   \right)\label{eq:equation43}
\end{equation}
The relevant Wolfenstein like approximation of the mixing matrix is reported in (\ref{eq:equation42a}).
\begin{table*}
\begin{multline}
\widehat{V}=\left( \begin{array}{cccc}
    C(\lambda)+\frac{A^2 \lambda^6}{4!}\Phi & S(\lambda)-\frac{A^2 \lambda^5}{2}(\Pi_1^*-\frac{1}{6})  \\ \\
    -S(\lambda)-\frac{A^2 \lambda^5}{2}(\Pi_1-\frac{5}{6}) &  C(\lambda)-(A\lambda^2)^2 \left[ \frac{1}{2}-\frac{\lambda^2}{3}\left( \frac{1}{4}-\imath \eta \right)\right]   \\ \\
        -A\lambda^3\left[ \left(\Pi_1-1\right)-\frac{\lambda^2}{6}\left( \Pi_1-\frac{3}{4} \right) \right] &
                         -S(A\lambda^2)+\frac{A \lambda^4}{4!} \left( \lambda^2(\Pi_1-\frac{7}{10})-12(\Pi_1- \frac{2}{3})  \right) \\ \\
    -\lambda^6 p (\Pi_2-1) & -\lambda^5 \left(- p+ A u \frac{\lambda}{2} \right)
     \end{array}
\right. \\ \\
\left. \begin{array}{cc}
     A\lambda^3\left[ \Pi^*_1-\frac{\lambda^2}{6}\left( \Pi^*_1 -\frac{1}{4} \right) \right] &    -\Pi^*_2 p \lambda^6\\ \\
    S(A\lambda^2) + \frac{A \lambda^4}{4!} \left( \lambda^2(\Pi^*_1-\frac{3}{10})-12(\Pi^*_1- \frac{1}{3}) \right) &\lambda^5 \left( p+ A u \frac{\lambda}{2} \right)    \\ \\
                         1-\frac{(A \lambda^2)^2}{2}+\frac{A^2 \lambda^6}{4!}\Phi  & u \lambda^4\\ \\
      -u \lambda^4 & 1
     \end{array}
\right)\label{eq:equation42a}
\end{multline}
\[
\begin{array}{l}
 \Phi_1=\rho_1+ \imath \eta_1 \;\; \mbox{with} \;\; \displaystyle{\rho_1= \frac{x}{y} \cos(\delta_1)+\frac{1}{2}, \;\; \eta_1=-\frac{x}{y} \sin(\delta_1)}; \;\;
\Phi_2=\rho_2+ \imath \eta_2 \;\; \mbox{with}\;\;\displaystyle{\rho_2=\frac{z}{p} \cos(\delta_2)+\frac{1}{2}, \;\; \eta_2=-\frac{z}{p} \sin(\delta_2)} \\
\Phi_j^*=\rho_j- \imath \eta_j  \;\; j=1,2
\end{array}
\]
\end{table*}

Furtheremore the invariants (4 for a $4\times4$ matrix), obtained directly from (\ref{eq:equation43}) read
\[
\begin{array}{rl}
J_2=&\lambda^2 f(1,y \lambda ,x \lambda^2)+\lambda^8 f(u,p\lambda,z\lambda^2) \\
& \mbox{where} \;\; f(a,b,c)=a^2+b^2+c^2  \\ \\
J_3=& 2\imath \lambda^6 \left[ xy \sin(\delta_1)+zp\lambda^6 \sin(\delta_2) \right]= \\
   =&xy\lambda^6 (e^{\imath\delta_1}-e^{-\imath\delta_1})+zp\lambda^{12}(e^{\imath\delta_2}-e^{-\imath\delta_2}) \\ \\
J_4=& u^2 \lambda^{10}+[ 2uyz\cos(\delta_2) -2pux  \cos(\delta_1) ]\lambda^{13} + \\
    &[p^2 x^2-px^2 y e^{\imath(\delta_2-\delta_1)}-pxyz e^{-\imath(\delta_2-\delta_1)} +xy^2 z]\lambda^{16}
\end{array}
\]
the first invariant, associated with the trace of A, is zero. \\
It is evident that the $J_2$ and $J_3$ invariants are just a generalization of those reported in
Eq. (\ref{eq:equation31}) while the fourth is completely new being associated to the full determinant of the matrix.
We have reported this example to show the flexibility of the method it is however evident that
the detection of CP violating effects due to the new phase require an
accuracy at least of the order $\lambda^6$. \\
Before concluding the paper we will address the problems associated with the exponential forms
of the neutrino mixing matrix, which have also been discussed in \cite{Dattoli3}, where the
lepton-quark complementarity \cite{Minakata} has been reformulated by noting that the relevant
rotation occur around axes forming an angle of $45^0$.
The present experimental data seem to favor the tribimaximal (TBM) form \cite{Harrison}
therefore the neutrino mixing matrix reads
\begin{equation}
\widehat{U}=\left(   \begin{array}{ccc}
        \displaystyle \sqrt{\frac{2}{3}}& \displaystyle \frac{1}{\sqrt{3}}  & 0 \\ \\
        \displaystyle -\frac{1}{\sqrt{6}}& \displaystyle \frac{1}{\sqrt{3}}& \displaystyle \frac{1}{\sqrt{2}}\\ \\
        \displaystyle \frac{1}{\sqrt{6}} &\displaystyle -\frac{1}{\sqrt{3}}& \displaystyle \frac{1}{\sqrt{2}}\\ \\
     \end{array}
\right) \label{eq:equation44}
\end{equation}
If we assume that also this form is generated by an exponential matrix
(with all real entries) according to
\begin{equation}
        \widehat{U}=e^{\widehat{B}}; \;\;\;
        \widehat{B}=\left(  \begin{array}{ccc}
                0\;\;& \;\;\alpha \;\;&\;\; \beta \\ \\
                -\alpha\;\; & \;\;0 \;\;& \;\;\gamma\\ \\
                -\beta\;\; &\;\;-\gamma\;\; &\;\; 0 \\ \\
            \end{array}
    \right)\label{eq:equation45}
\end{equation}
We obtain the following correspondence between the entries of the $\widehat{B}$ matrix
and those of the TBM form
\begin{equation}
\begin{array}{l}
        \widehat{B}=\alpha \left(  \begin{array}{ccc}
                0\;\;& \;\;1 \;\;&\;\; \displaystyle -\frac{1}{\sqrt{2}+1} \\ \\
                -1\;\; & \;\;0 \;\;& \;\;\frac{\sqrt{3}+\sqrt{2}}{\sqrt{2}+1}\\ \\
                \frac{1}{\sqrt{2}+1} \;\; &\;\;-\frac{\sqrt{3}+\sqrt{2}}{\sqrt{2}+1}\;\; &\;\; 0 \\ \\
            \end{array}
    \right) \\ \\
\alpha=\displaystyle {2\frac{\sqrt{2\sqrt{2}+3}}{\sqrt{2\sqrt{2}+2\sqrt{6}+9}} \frac{1}{\sin\left(
         \frac{\sqrt{3-\left( \sqrt{\frac{2}{3}}+\frac{1}{\sqrt{3}}+\frac{1}{\sqrt{2}}  \right)}}{2}\right)} }
\end{array} \label{eq:equation46}
\end{equation}
The values of the entries of the TBM matrix do not allow the interpretation of
the neutrino mixing matrix as an expansion around the unit, notwithstanding it
is possible to get a better agreement with experimental by making an appropriate
expansions around the matrix $\widehat{B}$ and then around the TBM, as it will be shown in a dedicated paper.\\
In this paper we have provided an extensive account of the possibilities offered by the exponential
form of the CKM matrix, which looks like a prototype from which all the other forms can be derived,
we hope that our suggestions provide a useful tool in the relevant applications.

\begin{acknowledgements}
The authors are deeply indebted to Dr. D. Babusci for stimulating discussions and comments during any stage of the paper.
\end{acknowledgements}

\end{document}